\documentclass[]{spie}  

\usepackage{amsmath,amsfonts,amssymb}
\usepackage{graphicx}
\usepackage[colorlinks=true, allcolors=blue]{hyperref}

\usepackage{color}

\title{Planet Formation Imager: Project Update}

\author[1]{John D. Monnier}
\author[2]{Michael Ireland}
\author[3]{Stefan Kraus}
\author[4]{Almudena Alonso-Herrero}
\author[5]{Amy Bonsor}
\author[6]{Fabien Baron}
\author[7,56]{Amelia Bayo}
\author[8]{Jean-Philippe Berger}
\author[9]{Tabetha Boyajian}
\author[10]{Andrea Chiavassa}
\author[11]{David Ciardi}
\author[12]{Michelle Creech-Eakman}
\author[13]{Willem-Jan de Wit}
\author[14]{Denis Defr\`ere}
\author[15]{Ruobing Dong}
\author[16,8]{Gaspard Duch\^ene}
\author[17]{Catherine Espaillat}
\author[13]{Alexandre Gallenne}
\author[18]{Poshak Gandhi}
\author[19]{Jean-Francois Gonzalez}
\author[5]{Chris Haniff}
\author[18]{Sebastian Hoenig}
\author[5]{John Ilee}
\author[20]{Andrea Isella}
\author[21]{Eric Jensen}
\author[5]{Attila Juhasz}
\author[22]{Stephen Kane}
\author[23]{Makoto Kishimoto}
\author[24]{Wilhelm Kley}
\author[5]{Quentin Kral}
\author[15]{Kaitlin Kratter}
\author[25]{Lucas Labadie}
\author[26]{Sylvestre Lacour}
\author[27]{Greg Laughlin}
\author[8]{Jean-Baptiste Le Bouquin}
\author[28]{Ernest Michael}
\author[5]{Farzana Meru}
\author[29]{Rafael Millan-Gabet}
\author[10]{Florentin Millour}
\author[30]{Stefano Minardi}
\author[10]{Alessandro Morbidelli}
\author[31]{Chris Mordasini}
\author[32]{Andreas Morlok}
\author[33]{Dave Mozurkewich}
\author[34]{Richard Nelson}
\author[7,56]{Johan Olofsson}
\author[35]{Rene Oudmaijer}
\author[36]{Chris Packham}
\author[13]{Claudia Paladini}
\author[35]{Olja Panic}
\author[10]{Romain Petrov}
\author[37]{Benjamin Pope}
\author[38]{Joerg-Uwe Pott}
\author[39]{Luis Henry Quiroga-Nuñez}
\author[40]{Cristina Ramos Almeida}
\author[41]{Sean N. Raymond}
\author[42]{Zsolt Regaly}
\author[1]{Mark Reynolds}
\author[43]{Stephen Ridgway}
\author[44]{Stephen Rinehart}
\author[7,56]{Matthias Schreiber}
\author[45]{Michael Smith}
\author[46]{Keivan Stassun}
\author[14]{Jean Surdej}
\author[6]{Theo ten Brummelaar}
\author[13]{Konrad Tristram}
\author[47]{Neal Turner}
\author[48]{Peter Tuthill}
\author[49]{Gerard van Belle}
\author[47]{Gautum Vasisht}
\author[2]{Alexander Wallace}
\author[50]{Gerd Weigelt}
\author[51]{Edward Wishnow}
\author[52]{Markus Wittkowski}
\author[53]{Sebastian Wolf}
\author[5]{John Young}
\author[54]{Ming Zhao}
\author[55]{Zhaohuan Zhu}
\author[7,56]{Sebasti\'an Z\'u\~niga-Fern\'andez}
\affil[1]{University of Michigan, United States}
\affil[2]{Research School of Astronomy and Astrophysics, Australian National University, Canberra, ACT 2611, Australia}
\affil[3]{University of Exeter, United Kingdom}
\affil[4]{Centro de Astrobiologia (CSIC-INTA), Spain}
\affil[5]{University of Cambridge, United Kingdom}
\affil[6]{Georgia State University, United States}
\affil[7]{University de Valparaíso, Chile}
\affil[8]{University Grenoble Alpes, France}
\affil[9]{Louisiana State University, United States}
\affil[10]{Observatoire de la C\^ote d'Azur, France}
\affil[11]{Infrared Processing and Analysis Ctr., United States}
\affil[12]{New Mexico Institute of Mining and Technology, United States}
\affil[13]{European Southern Observatory, Chile}
\affil[14]{University de Li\`ege, Belgium}
\affil[15]{The University of Arizona, United States}
\affil[16]{University of California, Berkeley, United States}
\affil[17]{Boston University, United States}
\affil[18]{University of Southampton, United Kingdom}
\affil[19]{University de Lyon, France}
\affil[20]{Rice University, United States}
\affil[21]{Swarthmore College, United States}
\affil[22]{University of California, Riverside, United States}
\affil[23]{Kyoto Sangyo University, Japan}
\affil[24]{Eberhard Karls University T\"ubingen, Germany}
\affil[25]{University zu K\"oln, Germany}
\affil[26]{Observatoire de Paris, France}
\affil[27]{Yale University, United States}
\affil[28]{University de Chile, Chile}
\affil[29]{GMTO Corp., United States}
\affil[30]{Institute of Applied Physics, Friedrich-Schiller-University Jena, Germany}
\affil[31]{University Bern, Switzerland}
\affil[32]{Westflische Wilhelms-University M\"unster, Germany}
\affil[33]{Seabrook Engineering, United States}
\affil[34]{Queen Mary University of London, United Kingdom}
\affil[35]{University of Leeds, United Kingdom}
\affil[36]{The University of Texas at San Antonio, United States}
\affil[37]{New York University, United States}
\affil[38]{Max-Planck-Institut f\"ur Astronomie, Germany}
\affil[39]{Leiden University, Netherlands}
\affil[40]{Instituto de Astrofísica de Canarias, Spain}
\affil[41]{Lab. d'Astrophysique de Bordeaux, France}
\affil[42]{Konkoly Observatory, Hungary}
\affil[43]{National Optical Astronomy Observatory, United States}
\affil[44]{NASA Goddard Space Flight Ctr., United States}
\affil[45]{University of Kent, United Kingdom}
\affil[46]{Vanderbilt University, United States}
\affil[47]{Jet Propulsion Lab., United States}
\affil[48]{The University of Sydney, Australia}
\affil[49]{Lowell Observatory, United States}
\affil[50]{Max-Planck-Institut f\"ur Radioastronomie, Germany}
\affil[51]{Space Sciences Lab., United States}
\affil[52]{European Southern Observatory, Germany}
\affil[53]{Christian-Albrechts-University zu Kiel, Germany}
\affil[54]{The New York Times Co., United States}
\affil[55]{University of Nevada, Las Vegas, United States}
\affil[56]{N\'ucleo Milenio Formaci\'on Planetaria (NPF), Chile}

\authorinfo{Further author information: (Send correspondence to J.D.M.)\\J.D.M.: E-mail: monnier@umich.edu}

\begin{document} 
\maketitle

\begin{abstract}
The Planet Formation Imager (PFI) is a near- and mid-infrared interferometer project with the driving science goal of imaging directly the key stages of planet formation, including the young proto-planets themselves.  Here, we will present an update on the work of the Science Working Group (SWG), including new simulations of dust structures during the assembly phase of planet formation and quantitative detection efficiencies for accreting and non-accreting young exoplanets as a function of mass and age.  
We use these results to motivate two reference PFI designs consisting of a) twelve 3\,m telescopes with a maximum baseline of 1.2\,km focused on young exoplanet imaging and b) twelve 8\,m telescopes optimized for a wider range of young exoplanets and protoplanetary disk imaging out to the 150\,K H$_2$O ice line.  Armed with $4\times8$\,m telescopes, the ESO/VLTI can already detect young exoplanets in principle and projects such as MATISSE, Hi-5 and Heimdallr are important PFI pathfinders to make this possible.  We also discuss the state of technology development needed to make PFI more affordable, including progress towards new designs for inexpensive, small field-of-view, large aperture telescopes and prospects for Cubesat-based space interferometry.
\end{abstract}

\keywords{Infrared interferometry, Planet formation, PFI}

\section{INTRODUCTION}
The Planet Formation Imager (PFI) project was initiated in 2013 following an interferometry workshop at the Observatoire de Haute Provence.  Over the years, the PFI Project (directed by John Monnier) has grown to include a Science Working Group (SWG, led by Stefan Kraus) and a Technical Working Group (TWG, led by Michael Ireland) involving around 100 scientists and engineers from around the world.  Various scientific and technology aspects of PFI have been explored already in a series of papers first at the 2014 SPIE
\cite{pfimonnier2014,pfikraus2014,pfiireland2014} and then at the 2016 SPIE
\cite{pfimonnier2016,pfikraus2016,pfiireland2016,pfiminardi2016,pfimozurkewich2016real,pfibesser2016,pfipetrov2016}.
More information on the PFI Project can be found at the project website \url{http://planetformationimager.org}.

\begin{figure} [ht]
   \begin{center}
   \begin{tabular}{c} 
   \includegraphics[height=4in]{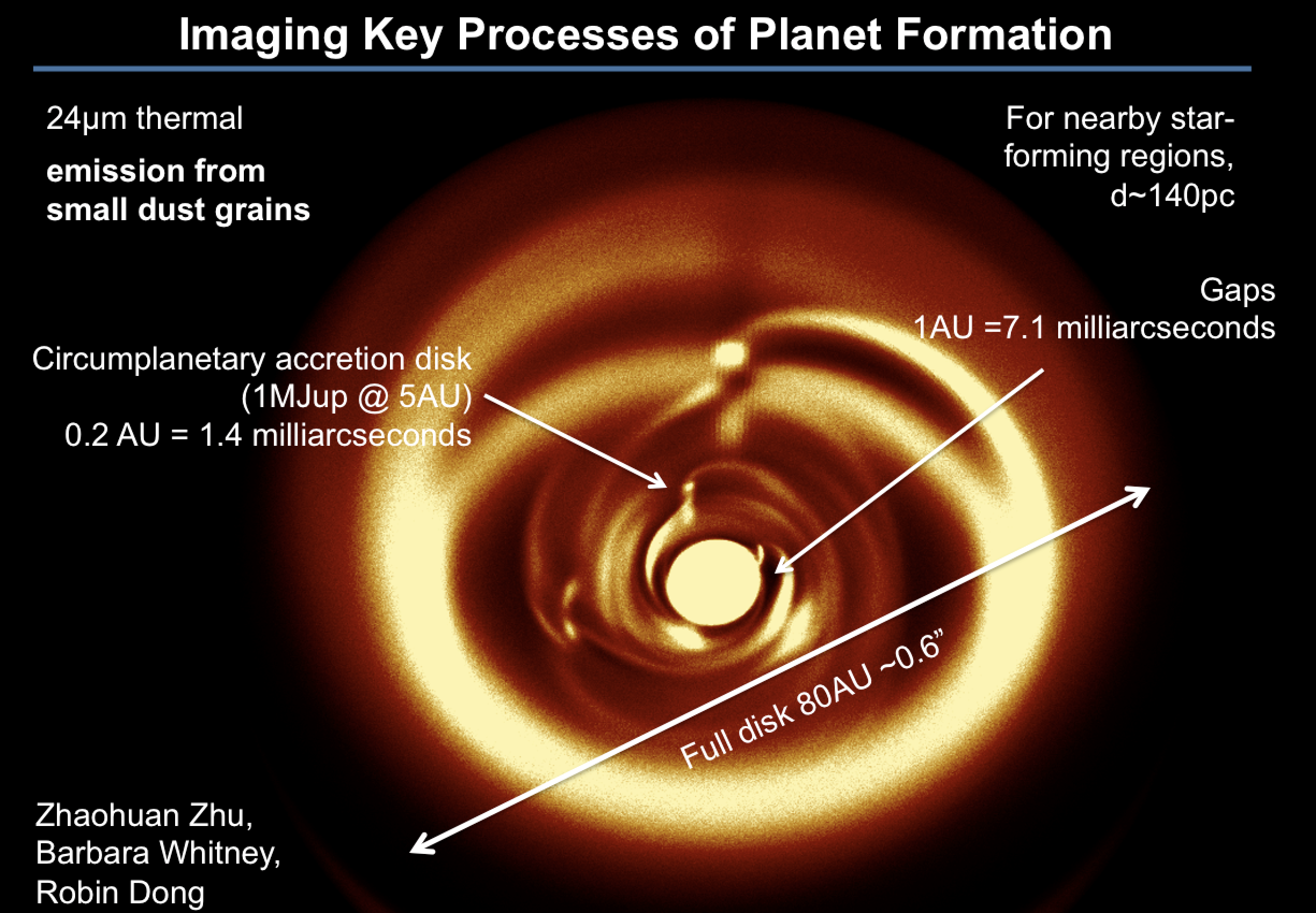}
   \end{tabular}
   \end{center}
   \caption[example] 
   { \label{fig:science} 
This figure shows the main planet formation processes that the Planet Formation Imager (PFI) Project wishes to address.  PFI should probe thermal dust emission to reveal gaps cleared by forming protoplanets (e.g., 1\,au gaps at 5\,au; 0.2\,au gaps at 1\,au) with enough resolution to just resolve circumplanetary disks themselves (0.2\,au for 1\,M$_{J}$ at 5\,au).  We also want to detect all the young and warm jovian planets themselves throughout the roughly 80\,au disk.  
   }
\end{figure}

Our founding scientific goals for PFI are to image key stages of planet formation {\em in situ} down to the scales of individual circum-{\em planetary} disks and with sensitivity to characterize young giant exoplanets themselves. With these bedrock goals, we then pursue more detailed science cases and seek technical solutions to achieve the required angular resolution and sensitivity.
The qualitative goals above have been quantified in a series of ``top-level science requirements'' TLSRs (see \S\ref{swg}).

The philosophy of the PFI project has been to develop the science goals first then see what facility can meet those goals. Earlier SPIE papers have explored the possibility of using ALMA, space interferometry, ELTs, visible interferometry and mid-IR interferometry for achieving the TLSRs.  For the most part, our team has preliminarily converged on a direct-detection 1.5-13\,$\mu$m (H,K,L,M,N bands) long-baseline interferometer on the ground to best achieve the goals, although heterodyne detection and space interferometry may offer other opportunities in the future as technology develops.  

We recognize that there needs to be a feedback loop between the SWG and the TWG. We are just now at the point in the project where we can directly compare the TLSR of the SWG to the achievable science from a given PFI facility architecture.  We expect to spend the new few years exploring if and how the new reference architectures put forward by the TWG can achieve the desired goals laid out by the SWG.  

In this brief report, we present an update on the activities of the PFI project and look toward the future.

\begin{table}[ht]
\caption{Typical absolute magnitudes for the emission components in protoplanetary disks \cite{pfimonnier2016,baraffe1998,spiegel2012,zhu2015,dong2015}, for the wavelength bands relevant for PFI, including adaptive optics system (Y band), fringe tracking system (H band), young exoplanets and dusty structures in the disk (L and N band).
To convert these absolute magnitudes to apparent magnitudes for an object located in Taurus at 140\,pc, simply add 5.7 magnitudes to the numbers below. }
\label{relevantfluxes}
\begin{center}       
\begin{tabular}{|l|c|c|c|c|}
\hline
Component & $M_Y$ & $M_H$ & $M_L$ &$M_N$ \\
          & (AO & (fringe & (dust & (dust \\
          & system)  & tracking) & \& planets) &  \& planets)\\
\hline
Example T Tauri Star & & & &  \\
\qquad 1 M$_{\odot}$, 2.1\,R$_{\odot}$, 3865\,K& 4.9 & 2.54 & $\sim$2.5  & $\sim$2.4 \\
\qquad 3\,Myr, [Fe/H]=0.0 &&&&\\
\hline
Protoplanet  & & & &  \\
\qquad ``hot start", 2M$_J$, 1\,Myr & & 12.9 & 11.0 & 9.1  \\
\qquad ``cold start", 2M$_J$, 1\,Myr& & 18.2 & 14.7 & 11.2   \\
\hline
Circumplanetary Disk & & & & \\
\qquad ($R_{in}=1.5~M_J$) & & & & \\
\qquad $ M\dot{M}=10^{-6}M^2_J\,{\rm yr}^{-1}$ & & 16.4 & 9.8 & 6.5  \\
\hline
4-planet gapped disk & & & &\\
\qquad Star only (2\,R$_{\odot}$, 4500\,K)  & 4.1 & 2.1 & 2.1  & 2.1  \\
\qquad Star + Disk (30$^\circ$ inclination) & 4.1 & 2.1 &  1.6 & -1.1  \\
\hline
\end{tabular}
\end{center}
\end{table}

\section{Update from the Science Working Group}
\label{swg}
\subsection{PFI Science Case}
The main job of the SWG is to keep the project science-focused and to establish the compelling top level science requirements.  
The SWG is working on an 11-paper special issue of Publication of the Astronomical Society of Australia (PASA). This ``Science Book'' will also contain description of one or two specific PFI facility architectures in order to allow a quantitative assessment of the achievability of the science goals. 

Figure~\ref{fig:science} gives a visual overview of the PFI science goals and relevant spatial scales.  We see PFI should have a field-of-view of at least 0.6'' to include the main portion of the planet-forming disk for nearby star forming regions. The angular resolution should be sufficient to not only resolve gaps caused by giant planet formation (e.g., 1\,au  gap at 5\,au) but also to resolve individual {\em circumplanetary} disks (e.g., 0.2\,au for 1\,M$_J$ at 5\,au). 

Given the excitement generated by recent imaging results from ALMA\cite{alma2015}, one might think all of these goals can be achieved with ALMA alone.  The HL~Tau image\cite{alma2015} showed a disk with multiple dark rings which could be due to interactions with a giant planet \cite{gonzalez2015} or possibly caused by other disk physics not involving a planet \cite{lorenaguilar2015}.  However, ALMA is not giving us a complete picture. For instance, the mm-wave emission is mainly showing large grains in the midplane of the disk and so much of the disk is invisible to ALMA.   Furthermore, ALMA has limited resolution of 25-75\,mas (depending on wavelength), corresponding to 3.5-10\,au at 140\,pc.  This is sufficient to just start to probe the region where giant planets are expected to form, but insufficient to see details. Note this resolution is not close to that needed explore gaps in disks within 5\,au (gaps of size 0.2-1\,au) or to resolve circumplanetary disks themselves.  There is some consideration about expanding baselines to up to factor 4 which would just bring the resolution limit to 1\,au, excellent to explore planet formation beyond the iceline but not enough to probe the Habitable Zone.  We have set a goal for PFI of imaging the dust around the 150\,K water iceline, as we expect this radius to be related to the zone of giant planet formation and includes the region of the disk where H$_2$O-rich asteroids form that eventually deliver water to terrestrial worlds\cite{raymond2017,walsh2011}.

The mechanism by which a young planet accretes dust and gas through circumplanetary disks is a poorly understood process and key ingredient to planet formation theory \cite{ayliffe2009}. The size scale of the disk is expected to be about $\frac{1}{3}$ size of the Hill Sphere $R_H=a\sqrt[3]{\frac{m_p}{3M_\ast}}$.  Molecules could be present in the region around an accreting planet as well as in the disk itself, such as HI (7-6), H$_2$O, CO, CO$_2$, CH$_4$, C$_2$H$_2$, NH$_3$ (e.g. Rigliaco et al.\ 2015\cite{rigliaco2015}). Also exoplanets might disturb the axisymmetric distribution of molecules in the disk, allowing for their discovery (e.g., as has been done recently with ALMA for very wide planets\cite{teague2018,pinte2018}).
Much more work\cite{ruge2014} is needed to understand how accreting protoplanets might be observable and with what tracers.

The last key science topic is to actually detect directly and characterize all the giant planets younger than 5-10\,Myr around young stars.  Giant planets have relatively high temperatures after they form\cite{spiegel2012} and are even brighter while accreting\cite{zhu2015}, making them ideal targets for high angular resolution searches.  While ELTs will be able to detect some far-out giant planets if the circumstellar dusty disk is gone (for reference, $\lambda/D$ at 3.5\,$\mu$m for ELT is 2.5\,au at Taurus, comparable to ALMA), an interferometer with kilometric baselines will have sufficient angular resolution to resolve out the dust to see young exoplanets even at the earliest stages.  We want to see where giant planets form and how they migrate or interact dynamically.  We expect to see significant differences in the location of giant planets at 1 versus 100\,Myr and PFI should be sensitive to stars at the young ages when the gas disk is still relevant to processes such as migration and before most dynamical instabilities have been triggered.  Indeed, understanding giant planet formation is key to understanding terrestrial planet formation\cite{raymond2006b,baruteau2014,davies2014}.

The SWG has collected some relevant fluxes of stars, young giant planets, and accreting protoplanets in Table~\ref{relevantfluxes}.  This information and the above science goals led to a set of ``top-level science requirements (TLSRs)'' and these are collected in Table~\ref{tab:tlsr}.

\begin{table}
\begin{center}
\caption{Top-level Science Requirements\label{tab:tlsr}
}
\begin{tabular}{|l|c|c|}
\hline
 Parameter & Dust Imaging & Young Exoplanets\\
\hline
Wavelengths & 5-13\,$\mu$m & 3-5\,$\mu$m \\
Typical Source Distance & 140\,pc & 50-500\,pc \\
Spatial Resolution & 2\,mas $\equiv$ 0.3\,au & 0.7\,mas $\equiv$ 0.1\,au (for 140pc)\\
Point-Source Sensitivity & $m_N\sim12.5$ (5-$\sigma$)  & $m_L\sim18.5$ (5-$\sigma$) \\
\qquad (t$=10^4$s) & & \\
Goal Surface Brightness (K) & 150\,K  &  $--$\\
\qquad (t$=10^4$s) & & \\
Spectral Resolving Power&&\\
\qquad Continuum      & R$>100$  & R$>100$ \\
\qquad Spectral Lines & R$>10^5$ & R$>10^5$ \\
Field-of-view &  $>0.15"$  & $>0.15"$ \\
Minimum Fringe Tracking Limit & $m_H<9$ (star only) & $m_H<9$ (star only)\\
Fringe tracking star & $\phi<0.15\,$mas & $\phi<0.15\,$mas \\
\hline
\end{tabular}
\label{tlsr}
\end{center}
\end{table}

\subsection{Highlights from Science Working Group}

\subsubsection{Number of targets}
The SWG has explored the number of YSO targets available in the sky as a function of brightness and disk class.  We estimate that approximately 4100 Class\,II sources brighter than K=12\,mag are observable for mid-latitude sites such as the ALMA site or the Anderson Mesa site near Flagstaff.  Having a large number of targets is essential for the exoplanet detection characterization science case since only $\sim$10\% of solar-mass stars have a giant planet beyond 0.5\,au\cite{cumming2008}.  By comparison, the YSO census of Class\,II objects that can be seen from Antarctica is only 350 sources brighter than K=12\,mag, strongly arguing against the High Antarctic Plateau as a site despite the potential breakthrough seeing conditions and low ambient temperatures\cite{lawrence2004}. More details will be contained in the PFI Science White Book to be published in PASA.

\subsubsection{Hydrodynamic Simulations of Planet-forming Disks}
\label{dust_simulation}

\begin{figure} [ht]
   \begin{center}
   \begin{tabular}{c} 
   \includegraphics[width=6in]{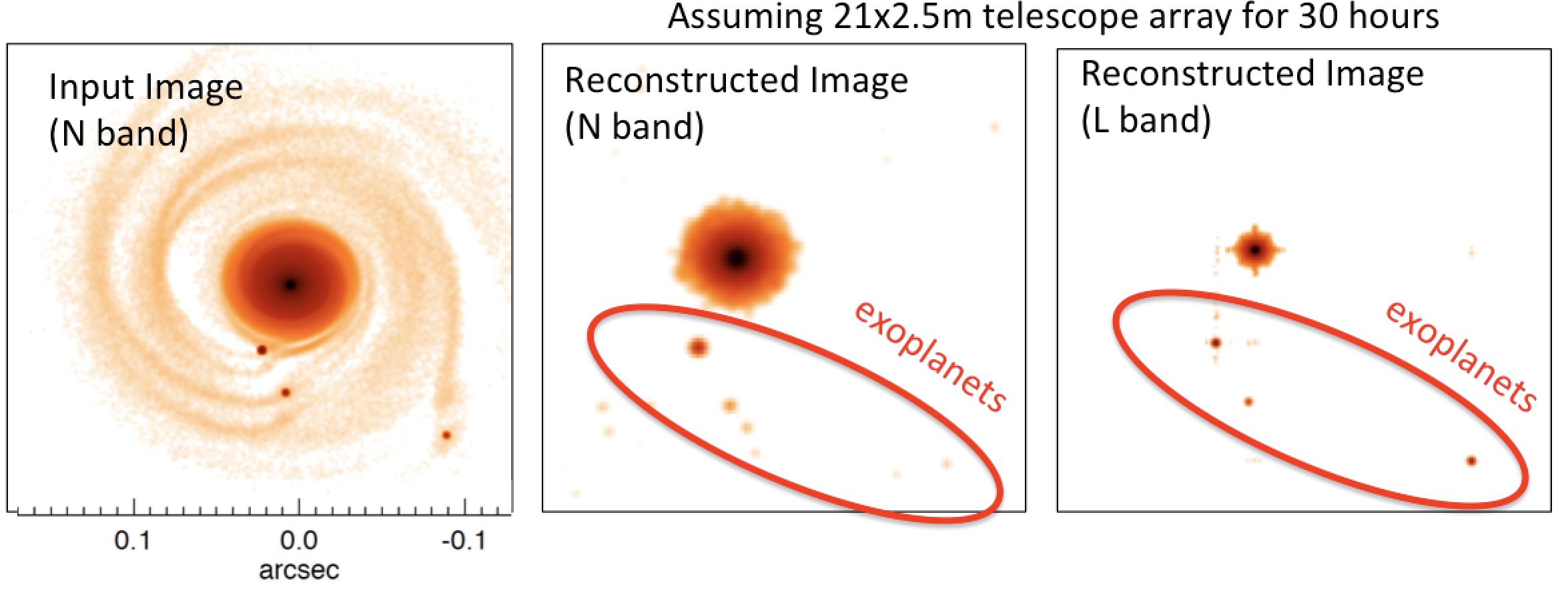}
   \end{tabular}
   \end{center}
   \caption[example] 
   { \label{fig:oldsim} 
This figure summarizes results of early PFI simulations from Monnier et al.\cite{pfimonnier2016}.  The left panel shows the surface brightness of a disk with 4 giant planets forming a transition disk based on a simulation by Dong et al.\cite{dong2015}.  The peak brightness temperature of the dust outside 5\,au was never above 100\,K and could not be detected in a simulated observation of $21\times2.5$\,m telescopes with a 3 hour integration in N~band (see middle panel) or L~band (see right panel), consistent with our analytical SNR calculations that pinned the sensitivity of this architecture to be $\sim$150\,K.   While imaging the cool dust was not possible, the warm emission from the young exoplanets themselves was clearly visible (see middle and right panels). 
   }
\end{figure}

In 2016, Monnier et al.\cite{pfimonnier2016} made a first attempt to reconstruct an image of a realistic planet-forming disk using a simulated PFI facility architecture of $21\times2.5$\,m telescopes.  Figure~\ref{fig:oldsim} summarizes those results.

We learned two major things from this first simulation.  
\begin{itemize}
\item We quantified that the surface brightness limit of 150\,K is not sufficient to image the optically-thin warm dust beyond the H$_2$O iceline in the first hydrodynamical simulation we scrutinized -- in fact, the brightness temperature never exceeded 100\,K thus staying orders of magnitude too faint to observe for a ground-based interferometer operating at 10\,$\mu$m.  We began to seek additional simulations over a wider range of assumptions and we present a new calculation in this paper.
\item The young forming exoplanets themselves were detected.  We were able to detect 3 of the 4 hypothetical exoplanets at both L and N~band using prescriptions from Zhu et al.\cite{zhu2015} for the active accretion phase and models from Speigel \& Burrows\cite{spiegel2012} for the emission from the gas giants themselves using ``hot-start'' models.  We have expanded our investigation into exoplanet detection in this paper in \S\ref{exoplanets}.
\end{itemize}

\subsubsection{New Dust Simulation}

The new model was based on the 1\,M$_J$ simulation of Dong, Zhu, \& Whitney\cite{dong2015} placed at 140\,pc and evolved for 0.4\,Myr. In brief,
\begin{itemize}
\item central star temperature 4500\,K, radius 2\,R$_\odot$, mass 1\,M$_\odot$, planet location at 5\,au, planet mass 1\,M$_J$, maximum disk diameter 80\,au, disk inner radius 0.094\,au (sublimation radius),

\item the initial temperature (scale height) and surface density profile can be found in section 2.1 of Dong, Zhu, \& Whitney\cite{dong2015}, namely temperature  $T=221 (r/{\rm au})^{-\frac{1}{2}}\,$K, scale height $h_g/r = 0.029 (r/{\rm au})^{0.25}$, gas surface density $\Sigma_g = 178 \frac{\rm au}{r} e^{-\frac{r}{100{\rm au}}}$ g cm$^{-1}$,

\item the small dust (well-coupled with the gas) are assumed to be interstellar medium dust (sub-micron-sized); the large dust are assumed to have a maximum size of 1\,mm and with a size distribution specified as equation (4) and table 1 (model 1) in Wood et al.\cite{wood2002}
\end{itemize}

\begin{figure} [ht]
   \begin{center}
   \begin{tabular}{c} 
   \includegraphics[width=6in]{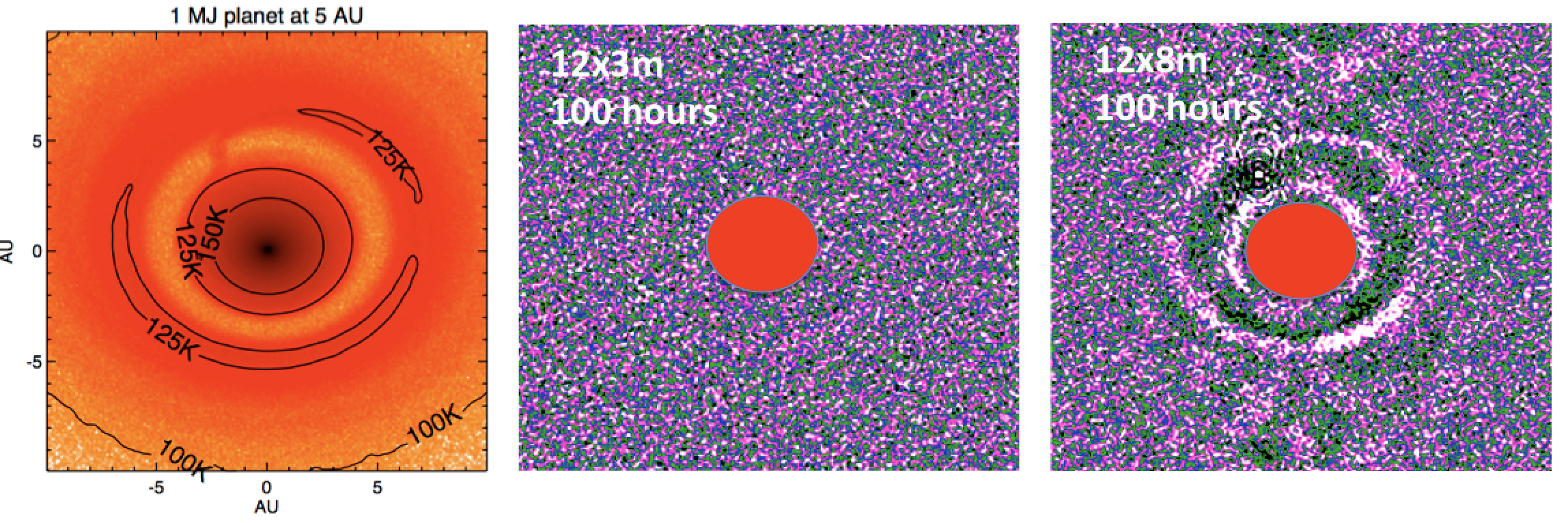}
   \end{tabular}
   \end{center}
   \caption[example] 
   { \label{fig:newsim} 
 This figure shows results from a new dust simulation of a single 1\,$M_J$ planet forming at 5\,au.  In the left panel, we see a mid-IR (10-11\,$\mu$m) image from a radiative transfer calculation based on work by Dong et al.\cite{dong2015}.  The peak brightness for the dust around the gap is T$_B\sim125$\,K, too faint to detect using simulated data from a $12\times3$m array configuration (see middle panel).  Sensitivity for background-limited observations scale like $\sqrt{N_{\rm tel}} D^2$ and we show a simulation for $12\times8$\,m telescopes in right panel with 7.1-times, or 2.1\,mag, greater sensitivity where dust structures around the gap are clearly visible (displayed using histogram equalization). See text for more details on these preliminary results.  The red ellipse shows the region of the disk that was nulled in this simulation (see text for more details). }
\end{figure}

Figure~\ref{fig:newsim} shows the results of our new simulation and our attempt to carry out aperture synthesis imaging reconstructions.  The left panel shows that the new simulation of a single Jovian planet has a higher brightness temperature (125\,K) at 5\,au than the original 4-planet simulation ($<$100\,K).  This is mainly due to the higher optical depth in the gap region when there is only a single planet.  We attempted to make a full image reconstruction using our PFI simulator and CLEAN deconvolution algorithm.  However, we found that our CLEAN algorithm was not advanced enough to suppress artifacts from the bright central region of the disks.  In order to proceed and see if the outer disk was present at high SNR in our simulated datasets, we ``null''ed out the center of the disk (i.e.\ removed it from the input data) and show here directly the ``dirty map'' based on gridding the full interferometric data set.  Note that in the mid-IR we do not expect to need to perform nulling since background noise swamps noise from the targets themselves (in most all cases) -- that said, care must be taken to have good calibration of data to achieve the necessary high dynamic range imaging to see faint dust emission at 5\,au just outside the bright inner disk.  The conclusion from this work is that the mid-IR surface brightness limit of a $12\times3$\,m PFI architecture is 150\,K, not enough to detect the gap in the disk formed by the simulated Jupiter.  However, the $12\times8$\,m PFI facility was was able to achieve 7-times higher SNR corresponding to $T_B=125$K, sufficient to just image the cool dust and gap surrounding the forming planet.

We note that even this simulation is likely pessimistic compared to the formation of our own Jupiter. The so-called ``grand tack'' model \cite{walsh11} has Jupiter forming at 3.5\,au, with higher gas densities that push the snow line to above 160\,K. Jupiter then migrates inwards to warmer disk regions prior to migrating outwards. In such a scenario, the cool dust gap is likely to be almost always visible at the design 150\,K brightness sensitivity for the 12$\times$3m PFI reference design. Extending our models to such a scenario will be part of our future work.

\subsection{Exoplanet Yields}
\label{exoplanets}

\begin{figure} [ht]
   \begin{center}
   \begin{tabular}{c} 
   \includegraphics[height=2in]{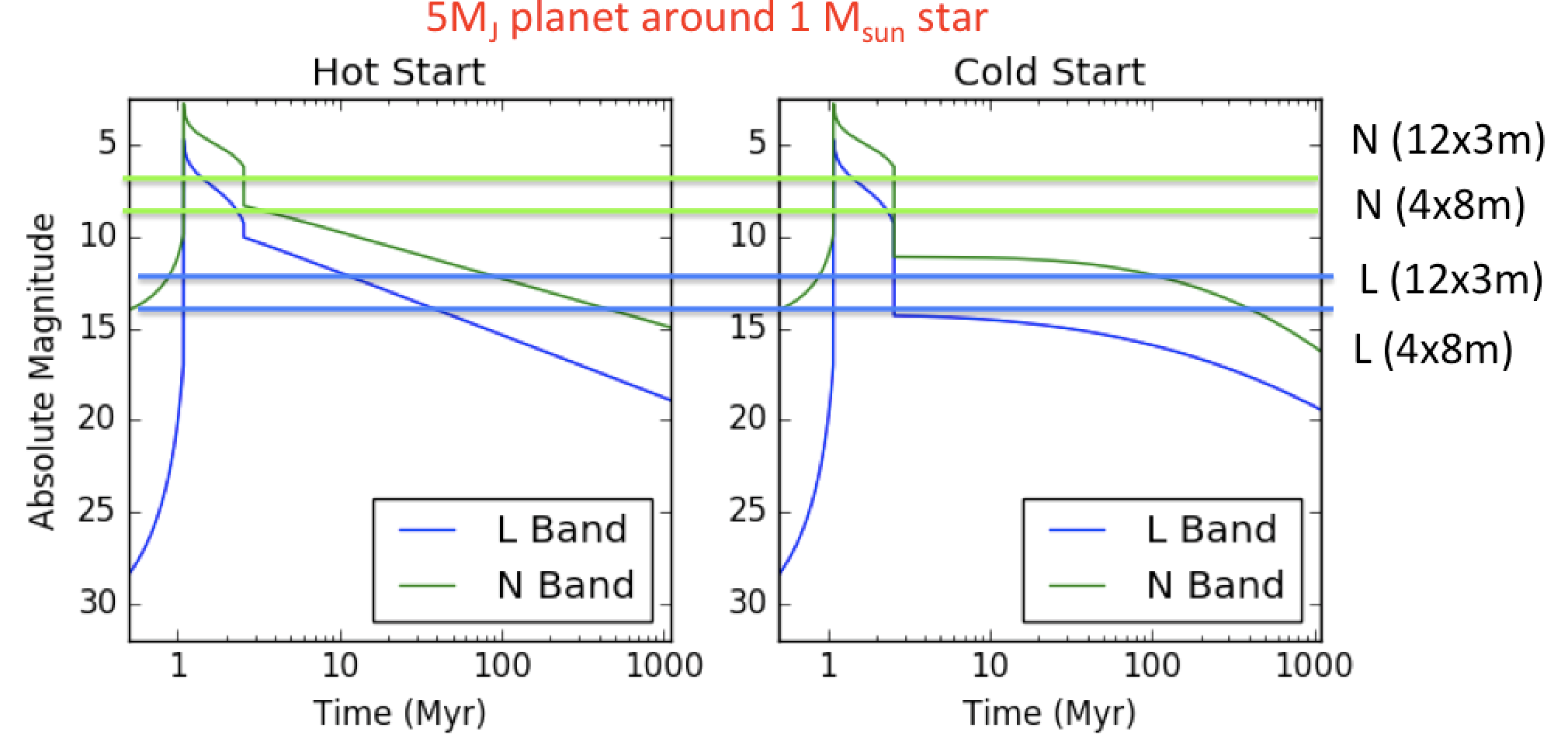}
      \includegraphics[height=2in]{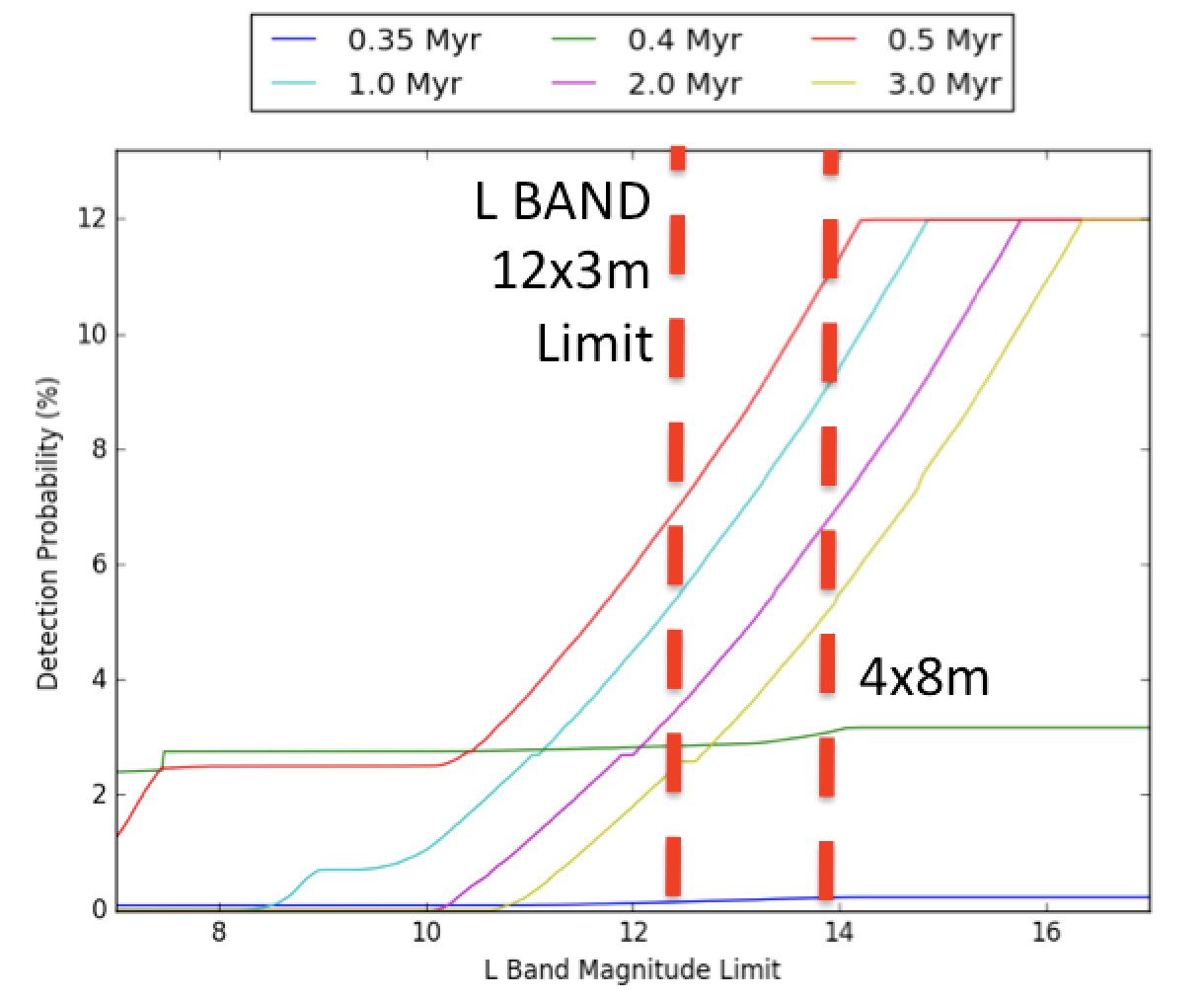}

   \end{tabular}
   \end{center}
   \caption[example] 
   { \label{fig:exo1} 
   Simulation results from Wallace et al.\ (2018, in preparation).  We include observing limits for a $12\times3$\,m array version of PFI and a $4\times8$\,m array resembling the ESO/VLTI with the Hi-5 instrument (Defrere et al., these proceedings).  (left, middle panels) These panels show the expected absolute magnitudes (L and N band) for a massive giant planet over time.  We see both $12\times3$\,m PFI and $4\times8$\,m Hi-5 can detect nearly all 5\,M$_J$ planets while actively accreting at both N and L band. L~band is better for detecting the planets themselves after the accretion phase.  (right panel) Here we see a preliminary estimate of what fraction of young stars will have an exoplanet of a given absolute magnitude in L band based on demographics of Cumming\cite{cumming2008}.  From this, we see that a $4\times8$\,m Hi-5 instrument is potentially more sensitive than a $12\times3$\,m PFI for this science case, here finding more than 2-times the exoplanets under 10\,Myr assuming hot-start\cite{spiegel2012} models and a Zhu\cite{zhu2015} prescription during active accretion phase (Assumptions: giant planets $M>0.5\,M_J$, orbit 0.25-25\,au).
}
\end{figure}

Figure~\ref{fig:exo1} (left and middle panels) shows the potential for exoplanet detection using $12\times3$\,m PFI or a future $4\times$8\,m Hi-5 instrument on the VLTI (Wallace et al 2018, in preparation).  For the 5\,M$_J$ planet considered here, we see that both architectures can detect all accreting\cite{zhu2015} exoplanets at both L and N bands while only hot-start models\cite{spiegel2012} are visible for L band.  The factor 4 (1.4mag) sensitivity advantage of the $4\times8$\,m Hi-5 design would allow many more planets to be discovered than the $12\times3$\,m PFI, especially if real planets follow cold-start models\cite{spiegel2012}.

Wallace et al. (2018, in preparation) also took the next step of calculating exoplanet yields based on our current understanding of exoplanet demographics around low-mass stars\cite{cumming2008}.  Figure~\ref{fig:exo1} (right panel) shows results by including giant planets $M>0.5\,M_J$ with orbits between 0.25-25\,au.
We see that $12\times3$\,m PFI would detect planets around 6\% of stars with planets $<$1\,Myr old while only detecting planets around 2\% of stars at age of 3\,Myr. The calculation also shows that (in principle) the $4\times8$\,m Hi-5 would see planets around $\sim 2$-times as many stars  assuming enough angular resolution to resolve the planets from the central stars.  This emphasizes that the Hi-5 instrument concept for $4\times8$\,m ESO/VLTI has potentially impressive sensitivity, although the relative short baselines of VLTI (max 120\,m) will only capture Jovian planets beyond about 1-2\,au, reducing the scientific reach compared to a PFI with kilometric baselines. The limited baseline lengths of VLTI will likely also hamper the detection of planets in systems with significant amounts of optically thick disk emission and we expect that PFI will quickly benefit from its larger number of apertures and longer baselines for the detection of planets in such systems. We also want to carry out this analysis at K band where thermal background is negligible and where nulling performance would be paramount.

One take away is that L band appears to be more sensitive band for exoplanet detection than N band assuming that nulling can be achieved to suppress emission from inner disk.  This is expected for young {\em warm} exoplanets since the planets have temperatures warmer than the IR background of a ground-based telescope (T$\sim$280\,K), thus wavelengths shorter than 10\,$\mu$m have improved SNR.  For older/cooler exoplanets, a space-based interferometer will be needed to detect characterize planets in general, except for the case when the ELTs can spatially resolve the planet from the star using coronagraphy.

\section{Update from the Technical Working Group}
\label{twg}
\subsection{Introduction}
The PFI SWG and TWG have published a large number of articles in SPIE, outlining both the science case \cite{pfikraus2014,pfikraus2016} and possible technical implementations of PFI \cite{pfimonnier2014,pfiireland2014,pfimonnier2016,pfiireland2016,pfiminardi2016,pfimozurkewich2016real,pfibesser2016,pfipetrov2016}.  This year there are about 13 contributions directly related to PFI:
\begin{itemize} 
\item  Mirror Development. Z\'u\~niga-Fern\'andez et al. 
\item  Fringe Tracking. Ireland et al., Folcher et al.
\item  Hi-5 nulling instrument for VLTI. Defrere et al.
\item  Mid-IR Photonics. Labadie et al., Tepper et al., Martin et al., Goldsmith et al., Lagadec et al.
\item  Heterodyne work. Michael et al., Besser, et al.
\item  Space Interferometry. Quanz et al.
\end{itemize}

In addition to this work, the PFI TWG will write at least one paper in the planned Science White Book to define a baseline PFI architecture to reference in the science chapters.

In this section, we will comment on the driving factors for PFI that control sensitivity, outline two possible reference PFI facility architectures, introduce the Technology Roadmap, and give some updates on technology development in recent years.

\subsection{Comments on Sensitivity}
\label{sensitivity}
Ireland et al.\cite{pfiireland2016} laid out the basic SNR equations for PFI that have been used in this paper for estimating our limiting magnitudes and sensitivity. For background-limited observations (i.e., not limited by systematics or nulling performance), a simplified version of the  SNR equation for point source detection is:
\begin{equation}
{\rm SNR}_{\rm pt} \propto \frac{\sqrt{N_t} D^2 \sqrt{t}}{\sqrt{B(T)}},
\end{equation}
where $N_t$ is then number of telescopes, $D$ is the telescope diameter, $B(T)$ is the background emissivity (if thermal, T is temperature).  Notice that the noise level is independent of the telescope size in the background limit since a diffraction-limited system has constant \'etendue ($\Delta\Omega A \sim \lambda^2$). 

Based on equation (1), the facility sensitivity grows steeply with aperture diameter ($\propto D^2$) and only weakly with number of telescopes ($\propto \sqrt{N_t}$).  As a dramatic real-world example, the $4\times8$\,m telescopes of the ESO/VLTI UT array has the same background-limited sensitivity for point source detection as an array of $200\times3$\,m telescopes!  Note that this conclusion is subject to some assumptions, as some novel beam combiners\cite{pedretti2000} can recover the SNR$\propto N_t$ scaling under certain conditions\footnote{SNR using Heterodyne detection also scales $\propto N_t$.}. Furthermore, the angular resolution and capability to do nulling depend on the geometry of the array and so an array of many 3\,m telescopes with long baselines designed for nulling will have superior capabilities in some ways than a $4\times8$\,m array with shorter baselines. Regardless of details, one sees that it is critical to push for large apertures for sensitivity and that it is not generally cost-effective to try to compensate aperture with more telescopes.  

One other issue to consider in designing an array is the surface brightness limit for mid-infrared observations.  Since the blackbody function peaks in the mid-IR for $T=300$\,K, we recognize that observing dust much cooler than this in young stellar objects will be fighting the exponential Wien's tail of the Planck function. Quantitatively, emission at $T=150$\,K is $\sim$1\% of the emission at $T=300K$ in the mid-IR.  Indeed, going further to $T=125$\,K is another factor of 7 in flux.  Based on this physics, it is not practical to observe thermal emission of dust much below $T<125$\,K in the mid-infrared at high angular resolution, especially from the ground.  Fortunately, the water iceline occurs at T$\sim$150\,K, opening up observing opportunities for PFI.

\subsection{Reference PFI Facility Architecture(s)}

The TWG is working to finalize reference facility architectures for use in designing detailed science cases.  We have adopted the facility characteristics for our reference architectures:
\begin{itemize}
\item 1.2\,km maximum baseline chosen to attain 0.2\,au mid-IR resolution at 140\,pc: a) to resolve planet-opened gap for Jupiter 1\,au, b) to resolve diameter of circumplanetary disk for exo-Jupiter (1\,M$_{J}$@5\,au)
\item $12\times3$\,m diameter array chosen to have $T=150$\,K 3$\sigma$ surface brightness in 10000sec: 
Sensitivity to dust at $T=150$\,K, the temperature for the water iceline for typical disks
\item $12\times8$\,m diameter array chosen for enhanced surface brightness limit ($T=125$\,K) to see gaps and dust structures for giants planets forming as far out as 5\,au, and an enhanced exoplanet yield.\footnote{During the PFI Community Meeting held during the 2018 SPIE meeting, a straw poll of $\sim$25 people in attendance overwhelmingly favored pursuing the $12\times8$\,m architecture over the $12\times3$\,m version based on the increased science capabilities.}
\item Sufficient fringe tracking margin (H band magnitude limit at least 13) to observe solar-type stars in nearby star forming regions, even with some extinction and visibility reduction by an inner disk.
\item Sufficient point source sensitivity to detect young giant exoplanets for a range of models.
\item Nuller design for exoplanet detections at K, L bands.  This will impact arrangement of telescopes within 1.2\,km baselines to optimize clever nulling schemes (e.g., Hsiao et al.\cite{hsiao2010}, Lacour et al.\cite{lacour2014}, Martinache \& Ireland\cite{martinache2018}).
\end{itemize}

Table\,\ref{pfiref} contains information on two reference facility architectures for PFI, one is a $12\times$3\,m array and the other a $12\times8$\,m array.  The main difference is a factor of 7 in sensitivity that is crucial for imaging warm dust at 5\,au and for a complete census of giant planets in young disks.  The notional construction costs contained in this table are based on estimates in Ireland et al.\cite{pfiireland2016} and with new pricing for $12\times$6.5\,m telescopes based on a possible scheme presented by Kingsley et al. in these proceedings (conference 10700, ``An inexpensive turn-key 6.5m observatory with customizing options'').

Note that only recently have we started to seriously consider the demands on the array geometry for nulling interferometry.  A future paper will need to address and model the impact on $uv$-coverage and maximum baseline when more short nulling baselines are included.  Furthermore, there is also a science case for imaging in polarized scattered light which will also require excellent nulling down at K band -- again, the impact on the array geometry and calibration precision are yet to be explored.

\begin{table}
\begin{center}
\caption{Technical Description of Reference PFI Architectures\label{pfiref}}
\begin{tabular}{|l|c|c|}
\hline
Parameter & 12$\times$3m PFI & 12$\times$8m PFI \\
\hline 
Number of Telescopes & 12  & 12 \\
Telescope Diameter & 3\,m  & 8\,m \\
Maximum Baseline & 1.2\,km & 1.2\,km \\
Goal Science Wavelengths & 3--13\,$\mu$m  &3--13\,$\mu$m \\
Fringe-tracking wavelengths & 1.5--2.4\,$\mu$m & 1.5--2.4\,$\mu$m \\
Fringe tracking limits (point source) & $m_H<$13  & $m_H<$15 (AO-dependent)  \\
Point source Sensitivity (10$^4$s) & 18.1 (L), 12.2 (N) & 20.2 (L), 14.3 (N)   \\
Surface Brightness Limit (10$^4$s,$B=1.2$km) & 150\,K (N)& 125K (N) \\
Field-of-view & 0.25" (L), 0.7" (N) & 0.09" (L), 0.25" (N)\\
Note & w/ Nulling (2-4\,$\mu$m) &  w/ Nulling (2-4\,$\mu$m) \\
Construction cost & \$250M & \$600M$^*$ \\
\hline
\end{tabular}
\label{pfispecs}
\end{center}
{* Telescope cost based on informal estimate for 12$\times$6.5m telescopes based on Kingsley et al. (these proceedings)}
\end{table}

\subsection{Current Technology Roadmap and recent progress}
Table~\ref{tab:roadmap} contains a list of technologies that the TWG has identified as strategic and key investments for PFI.  Next, we collect some highlights of recent work in these areas.

\begin{table}[ht]
\caption{PFI Technology Roadmap} 
\label{tab:roadmap}
\begin{center}       
\begin{tabular}{|l|l|}
\hline
Critical Technology &  Considerations \\
\hline
Inexpensive telescopes & Possible key technologies: \\ 
& Replicated parabolic lightweight mirrors,\\
& inexpensive primaries with high low-order errors,\\
& lightweight structures with exquisite AO controls, \\
& {\bf Partner with industry, engineers}\\
& {\bf New telescopes for existing arrays}\\
\hline
L/M band IO combiners 
& Needed for high precision calibration, \\
& Explore Chalcogenide integrated optics. \\
& {\bf Pilot project at VLTI such as Hi-5 } \\
\hline
Wavelength-bootstrapped  
& L band imaging require $10^6$:$1$ dynamic range imaging, \\
fringe tracking 
& ultra-accurate fringe tracking in L based on H-band, \\
 & {\bf VLTI/GRA4MAT mode}\\
& while maintaining very high sensitivity, \\
& {\bf New ``high sensitivity''  fringe tracker at VLTI}\\
\hline
Mid-IR laser comb heterodyne & Possible ``add-on'' to L/M band \\
 & {\bf Develop combs, detectors, digital processing} \\
\hline
Low-cost operations model  & New array of limited scope \\
& {\bf e.g., {\em Gaia}/TESS follow-up interferometer} \\
\hline
Space interferometry & Longer-term future for high sensitivity \\
& Demonstrate formation flying with Cubesats\\
& {\bf Support new space missions; advocate in Decadal Surveys} \\
\hline
\end{tabular}
\end{center}
\end{table}

\subsubsection{PFI Telescopes}
One of the key hurdles to surmount in developing PFI is economic.  At the current prices for `traditional' telescopes\cite{vanbelle2004} scaling as D$^{2.5}$ -- roughly \$5M for a 2-meter aperture, $\sim$\$40M for 4-meter apertures -- an array of a dozen or more large telescopes rapidly becomes cost-prohibitive.  As such, one technical development area of the PFI project has been for demonstrating order-of-magnitude cost reductions in telescope construction.  

\noindent {\bf Adapting crude primary mirrors using deformable secondaries}
(lead: Gerard van Belle)

Advances in deformable mirror (DM) technology - and cost - mean that large but 'crude' (by historic standards) optics can be used for light collection, and corrected with a large-stroke DM system.  As long as the wavefront error of the large optics is large amplitude but constrained to low spatial frequencies (e.g. smooth, with no surface roughness scattering light), a DM matched to the large optics can recover the wavefront.  Taking advantage of such a design, deliberate amounts of sag can be designed into the large optics as well, simplifying the structural design of the optics.  

This design approach has been prompted by two observations: first, the scale of wavefront errors induced by atmospheric turbulence is similar to the wavefront errors found in off-the-shelf float glass, at $\sim$1-10\,$\mu$m.  Given that we have DM systems that are capable to accommodate atmospheric errors at this amplitude - at significantly faster correction rates - it is expected these systems can trivially accommodate large optics wavefront errors.  Second, the existence proof of correction of large-but-smooth errors -- namely, the Hubble primary -- indicate that this approach will be valid.

An additional significant simplifying factor that will enable this approach is the fact that  PFI apertures need to be diffraction limited, but effectively only on-axis.  This enables further design simplifications (e.g. DM-corrected spherical optics).  Towards the end of validating this concept, we are in the process of pursuing a lab demonstration of DM-corrected float glass apertures.

\noindent {\bf Spherical primaries with aspheric secondary correctors}
(lead: Amelia Bayo)

The PFI concept offers interesting challenges in the near future, one of them being the efficient manufacturing of $\sim$20 or more NIR-MIR telescopes. Traditionally Chile has been a host of the state-of-the-art observatories but has contributed little to the technological developments behind these facilities. In Valparaiso, in a collaboration between astronomers, particle physicists and engineers, we are trying to tackle the challenge (and change the paradigm in the Chilean tradition), and explore two possibilities for cost-effective and easy replicability astronomy mirror production. The project is very young (just a few months) and explores the possibilities of both, glass-based mirrors and carbon fiber ones. 

So far, the team has put together a small optical lab at Universidad T\'ecnica Federico Santa Mar\'ia, and has secured the funds for larger equipment that will allow to pass from the few tenths of centimeter prototypes in glass, to our goal for the next 2.5 years that is a one meter prototype. On the other hand, the work on the carbon fiber prototype has allowed us to identify a weakness in the accuracy of the replicability that we are in the process of solving by using an autoclave to increase the pressure in the copying procedure.
  \begin{figure} [ht]
   \begin{center}
   \begin{tabular}{c} 
   \includegraphics[height=5cm]{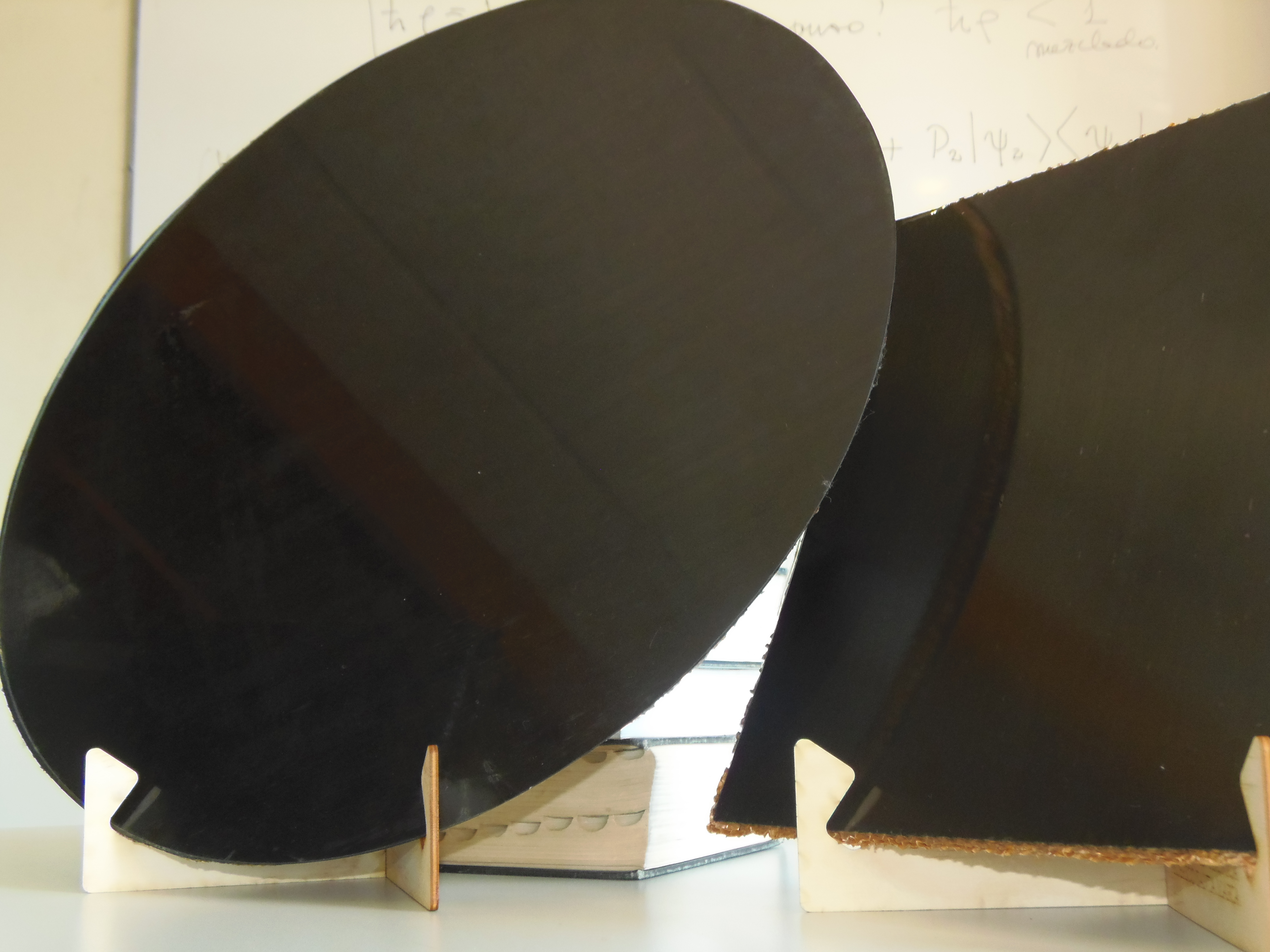}
   \end{tabular}
   \end{center}
   \caption[example] 
   { \label{fig:bayo} 
Spherical curvature carbon fiber prototypes (no coating has been applied to these prototypes) structured via an aramid honeycomb core. The shaping was obtained using the 50\,cm stainless steel mandrel in this case. (Credit: Bayo)}
   \end{figure}

\subsubsection{Heterodyne Developments}

{\bf Mid-IR laser combs and high-speed detector arrays} (lead: Gautum Vasisht)

Gautum Vasisht has reported on work at JPL and Caltech to develop mid-IR laser combs and the needed detector arrays.   Recent preliminary results include: 
\begin{itemize}
\item Directly generated 9.85\,$\mu$m combs with 60-80 lines with intermode spacing of 15 GHz with a total potential bandwidth approaching 1 THz. Power per line is about 1 mW.
\item Testing R-QWIP mixer technology in collaboration with Ed Wishnow's team (Berkeley SSL). Our QWIP samples are 25~pixel linear arrays with 30-70 \% QE. 
\item Tested MgB2 hot electron bolometers as 10\,$\mu$m mixers at JPL.  Bandwidth appears to be larger than 6GHz with devices not yet optimized for 10\,$\mu$m. 
\item Frequency generation experiments at Caltech using GaSe crystals not achieving high long-wave power so far.
\end{itemize}

\noindent {\bf Heterodyne Laboratory Testing} (lead: Michael Ernest)

Ernest Michael (Chile) is leading a lab-effort to maximize the SNR of heterodyne receivers. He is reporting work on a  novel prototype of a balanced detector correlation receiver at 1550\,nm wavelength (economic components in the telecom band) which was tested in
the Astro-Photonics Laboratory (APL) at the University of Chile
engineering campus in conjunction with a digital correlator. Each of
the two balanced receivers performed with a sensitivity equivalent to
a noise temperature of 2-2.5 times above the quantum limit (QL) using
the auto-correlation outputs of the correlator (contribution 10701-51
to this conference). This group reports that the cross-correlation output of these balanced
receivers, however, performs with a sensitivity equivalent to a noise
temperature a factor of 10-20 below the noise temperatures of the two
single-telescope balanced receivers. 
Assuming that the first results for a ``balanced correlation receiver''
described above will be corroborated, this makes a heterodyne approach
for the M, N- and L-bands very appealing if we consider that new
technology to broaden the optical receiver bandwidth comes
increasingly into reach\cite{pfiireland2014}, using a compact photonic
technique for the dispersion, like chip-integrated ring-resonators.
These separated spectral intervals are coupled one-by-one to multiple
(arrayed) balanced heterodyne detectors. Here it is to be considered
supportive that 25 GHz-class digital correlators will be soon
available. The APL seeks to demonstrate such a technique first at 1550\,nm, and test it at a 2-3 small-telescope heterodyne interferometer
prototype they currently develop (contribution 10701-94 to this
conference). Later this could be extended towards the mid-infrared
using the results at JPL/Caltech. Simulations of balanced receivers for
mid-infrared wavelengths are being performed currently at the APL by a
new postdoctoral semiconductor-devices simulation-expert.

\subsection{Focus on intermediate instruments and facilities}
Over the past two years, a few new proposals for instruments and facilities have been developed that could act as pathfinder for PFI.  Here we introduce a few.

\subsubsection{Hi-5 and Heimdallr}

The Hi-5 project is a science-driven international initiative to develop a new VLTI instrument optimized for high dynamic range observations \cite{Defrere:2018}. The current dynamic range of VLTI instruments (approximately 1:500) limits the ability to address some particularly exciting science cases and higher dynamic ranges (1:1000 to 1:10,000) have been demonstrated by nulling interferometers in the northern hemisphere and non-redundant aperture masking experiments. Reaching a similar or even higher dynamic range in the thermal near-infrared (L and M bands) would be extremely valuable to directly detect forming and young giant exoplanets, especially in the Southern hemisphere where most young stellar clusters are located. Such observations would be an important explorative program for PFI and contribute to its prime target list.

Hi-5 and PFI also share many common technology challenges, for instance on thermal near-infrared beam combination, accurate/robust fringe tracking, and nulling schemes. A wealth of instrumentation developments are currently underway to build high-contrast beam combiners in the L- and M-bands using single-mode fibers and integrated-optics components. New promising beam combination strategies are also under investigations, such as ``kernel nulling’’, which combines the advantages of nulling and closure phase and provides a metric robust against imperfect cophasing of the incoming stellar light \cite{martinache2018}.

Heimdallr is a proposed fringe tracker (PI: Michael Ireland) to enable high precision and high sensitivity fringe tracking at VLTI.  Heimdallr could be used with Hi-5, using the kernel-nulling (VIKiNG) architecture\cite{martinache2018} to obtain very well-calibrated nulling interferometry. 

Note that MATISSE is a L,M,N band 4-telescope combiner for VLTI that was just commissioned (Lopez et al., these proceedings).  MATISSE should be able to be used with GRAVITY in fringe-tracking mode (GRA4MAT) and reach very high sensitivity, although with lower resolution than the main PFI science goals.
Hi-5 and Heimdallr will build on great achievements expected from MATISSE and GRA4MAT, focusing on nulling and fainter fringe-tracking limits.
 
We showed in \S\ref{sensitivity} that the ESO/VLTI $4\times8$\,m array should have equivalent sensitivity to a $200\times3$\,m array (!).  While VLTI has relatively short baselines, the sensitivity is unrivaled -- this could be an amazing testbed for future PFI technology developments and science instruments. The PFI-TWG strongly advocates developing the VLTI-UT interferometric capabilities during the 2020s.

\subsubsection{{\em Gaia}/TESS follow-up: a new $3\times 3$\,m facility with 1\,km baselines}
While exoplanet science and thermal dust imaging of cool dust requires large aperture telescopes, a small array of $3\times3$\,m telescopes with $\sim$1\,km baselines could test key PFI technologies such as cheap telescopes, long delay lines, and low-budget operations model while achieving impressive science goals. In particular, {\em Gaia} and TESS will identify thousands of nearby binary stars and red giants that need following up. By measuring the separation of {\em Gaia} binaries with photocenter orbits, a $3\times3$\,m interferometer would measure thousands of new masses for a wide range of stars in the solar neighborhood. Furthermore, diameters of TESS red giants with astereoseismic radii from precision photometry will allow independent calibration of this key technique.

\subsubsection{Provocative Directions}
A few other provocative directions for the PFI project are worth considering:
\begin{itemize}
\item Adapt existing 300\,m-baseline sites (NPOI, CHARA, MROI) to use 3+\,m size telescopes and pursue L/M band exoplanets with nulling.
\item Adding new 8\,m telescopes around the four existing 8\,m telescopes at the ESO-Paranal site, building toward a $12\times8$\,m array eventually.
\item Develop space interferometry aggressively, including a coordinated international effort to debut formation flying.
\end{itemize}

\section{Looking forward}

The key science working group development will be to continually refine the critical surface brightness and imaging complexity requirement in the context of giant planet formation and migration theories. Also, further theory work is needed to obtain realistic models for spectral line tracers that will be used to determine the physical conditions and kinematics of the circumplanetary disk. As the array architecture takes shape, it will also be timely to broaden the science scope and to explore secondary science cases systematically, for instance in extragalactic astronomy.

We see the technical development of PFI as unfolding over the next decade with multiple pathfinder experiments along the way. For instance, the $4\times8$\,m telescopes of the VLTI should be fully exploited to demonstrate the $\sqrt{N}D^2$ sensitivity gains predicted by theory, to characterize young exoplanets around nearby field stars, and to determine dominant modes of planet formation through observations of planet-forming disks with their warm accreting protoplanets.
This will involve GRA4MAT fringe tracking for MATISSE and development of a new nulling instrument such as Hi-5 with high-performance fringe tracking (e.g., Heimdallr).

We see a further need to develop new technologies for large aperture telescopes and kilometric baselines. A $3\times3$\,m array spread over 1\,km could fill critical gaps in our galactic model by measuring masses for {\em Gaia} binaries and calibrate asteroseismic measures from TESS giants using stellar diameters.

Lastly, space interferometry has the potential to extend ``PFI science'' to the low-mass (possibly Earth-mass) planetary mass regime, and through cooled apertures has the potential to fully explore the 5--20\,au regime where disk gaps remain resolution-limited for ALMA and sensitivity-limited for a ground-based PFI.  The PFI-TWG strongly supports taking some steps in the next few years to launch a Cubesat interferometer, to explore Cubesat formation flying missions,  and to participate in NASA/ESA planning processes to explain the massive advantages of developing interferometric space capabilities for missions from far-infrared to x-rays.

In addition, the PFI community should become involved in other techniques that will expand and develop ``PFI science'', such as with extreme adaptive optics on the ELTs and longer baselines for ALMA, or with novel approaches such as highly-multiplexed heterodyne interferometry.



\bibliography{JDM_SPIE2018,pfi1,JDM_SPIE2016} 
\bibliographystyle{spiebib} 

\end{document}